\documentclass[twocolumn,aps,prl,groupedaddress,amssymb,showpacs]{revtex4}
\usepackage{graphicx}
\usepackage{dcolumn}
\usepackage{amsmath}
\usepackage{comment}
\usepackage{bm}

\begin{document}
\title{True wavefunctions and antiparticles of Klein-Gordon equation: 

$\hbar$-conjugation, elimination of $CPT$-invariance, and anti-gravitation}
\author{A. E. Kaplan$^{ \dagger }$ }
\email{alexander.kaplan@jhu.edu}
\affiliation{$^{\dagger}$The Johns Hopkins University, Baltimore, MD 21218, 
United States of America} 

\date{August 20, 2018}
\begin{abstract}
While revisiting Klein-Gordon relativistic
quantum equation for spin-0 particles,
we predicted that $\hbar$ reverses  its sign
for negative energies,
and formulated a universal symmetry rule,
whereby all the parameters that couple particles
to external fields reverse their sign along with $\hbar$
at a particle$\leftrightarrow$antiparticle 
transformation;
this in particular implies anti-gravitation
between matter and antimatter.
Our results suggest that
the $\hbar$-conjugation principle
and related invariance 
may replace $CPT$-invariance in general
relativistic quantum mechanics.
\end{abstract}
\maketitle
Spin-0 particles (e. g. Higgs-boson)
are arguably the most fundamental 
objects of relativistic quantum mechanics,
while described by the most 
basic,  Klein-Gordon (KG), equation [1]
(including composite particles [2]).
This makes KG-equation a good ground to
demonstrate a stunningly unexpected new invariance
toward sign-reversal of both the Planck constant $\hbar$
and all "charges" (including
electrical charge and gravitation mass)
in a particle$\leftrightarrow$antiparticle 
transformation, replacing thus a standard 
$CPT$-invariance for such a process. 

Compared to the Schr\"{o}dinger equation
which is a workhorse of
non-relativistic quantum mechanics,
the major relativity-related feature
of KG-equation [see eq. (2) below]
was the prediction of
negative energies and hence - antiparticles.
However, it was exactly that feature 
that made KG-equation plagued with many thorny issues,
such as e. g. negative probabilities for antiparticles;
even for "regular" particles,
the problem was that a "full" probability 
to find a particle in the space becomes Lorentz-$non$invariant;
a related (albeit apparently
less known, see below)
problem is a wrong amplitude of
the wave function for the  Doppler effect
pointed out by us here.
There were many attempts to fix the problem,
yet it is often believed by now that KG-equation
(and its Lorentz-invariant solution, see below eq. (3))
create unsurmounting difficulties for real problems [3].
As a result, when addressing antiparticles,
for example, most of modern day work
based on wave equations
largely rely on Dirac's equation [4] and
interpretation, rendering
KG-antiparticles essentially irrelevant [3-5].
Furthermore, within last few
decades the paradigm in the relativistic
quantum theory is shifting
toward the quantum field theory (QFT) [6],
bypassing thus quantum waves equations,
including KG-equation.

In this paper, we revisit
the KG-equation and its solutions
to find a universal fix that solves all
its problems/controversies, and
makes it a well-tuned
theoretical tool, and then turn to 
particle$\leftrightarrow$antiparticle
transformation and its new invariance.
The question is then, why do it at all
if QFT can supposedly handle those controversies? 
The first part of the answer is obvious:
in the same way that we need and daily use a regular 
Schr\"{o}dinger equation for non-relativistic QM-problem
in tremendous number of areas of physics and chemistry,
without revoking respective QFT, 
we need the KG-equation for
"regular" relativistic QM without antiparticles 
(e. g. to avoid Doppler-paradox, and to
have Lorentz-invariant probabilities).
When the spin of particles can be neglected,
the KG-equation serves as a good approximation too.
This greatly broadens its application
to the development of consistent quantum 
theory of relativistic electrons
in such systems and effects as e. g.
undulators and free-electron lasers [7],
relativistic hysteresises 
in cyclotron resonances [8] 
and related higher-order 
harmonics and subharmonics [9].
A good testing ground for those applications
and solution (4)
(both from  theoretical and experimental points of view)
could be so called "intermode traces" or
"quantum carpets" [10]
in a weakly relativistic case [11]
that may provide a sensitive
test by measuring phases of eigen-modes
with respect to each other.
To apply QFT instead of KG wave equation
to all those problems
would be akin to using QM in baseball game.

The second part of the answer
is that within our treatment
based on KG-equation we obtain
tantalizing new results for
antiparticles ($\hbar$-conjugation, 
new invariance that replaces $CPT$-invariance
and predicts anti-gravity between particles
and antiparticles),
which suggest that similar features can 
be expected with great probability in respective QFT,
so it would be of great importance to try and
generalize/translate them into full-blown QFT.
In fact, the next step in our 
planned research is exactly this -
to search and explore similar features 
in QFT description of bosons.

A standard derivation of KG-equation 
is based on (a) Einstein formula
for energy $ {E} $ vs. momentum $\vec{p}$,
$ {E}^2 =$ $ ( p c )^2 +$ $ ( m_0 c^2 )^2 $,
where $m_0 = inv > 0$ is an 
inertial rest mass of a particle,
(b) standard quantum operator recipes 
\begin{equation}
\hat{ {E} } = i \hbar   ( \partial / \partial t ) ,  \ \ \    
\hat{{ \vec{p} }} = - i \hbar \vec{\nabla} ;    \ \ \ with \ \ \
\hat{{\tilde{p}}} = \big( \hat{ {E} } / c , \hat{{ \vec{p} }} \ \big)
\tag{1}
\end{equation}
where $\hat{{\tilde{p}}} = i \hbar\hat{{\tilde{\mathcal{D}}}}$
is 4-momentum and ${\tilde{\mathcal{D}}}$ is
4-gradient operators 
(here and anywhere in this paper "tilde" denotes 4-object),
and (c) Lorentz(L)-invariant D'Alambert operator 
$\Box = ( \partial / c \partial t )^2  -  \vec\nabla^2$, where  
$\vec\nabla^2$ is a Laplacian.
The above Einstein relationship in operator form, 
applied to a wavefunction $\Psi$, yields then the KG equation:
\begin{equation}
\big[ \ \hbar^2 \Box + ( m_0 c )^2 \ \big]  \Psi = 0  .
\tag{2}
\end{equation}
In our theory, we first need
to fix well known problems/controversies
with eq (2) -- or, to be precise, with 
the choice of its solution.
To do so, we start from noting that
a crucial KG-stipulation historically
[1,2] (and wrongly, see below) 
imposed on scalar $\Psi$ is 
that it must be true (or 4-)scalar,
i. e. be $L-invariant$,
\begin{equation}
\tilde{\Psi} '  ( t',  \vec{r} ' ) =  
\tilde{\Psi}  ( t , \vec{r} )  =
\tilde{\Psi} {\bf \big[} \ t ( t' , \vec{r} ' ) , \ 
\vec{r} ( t' , \vec{r} ' ) \ {\bf \big]} 
\tag{3}
\end{equation}
where "prime" denotes functions 
or variables in a moving frame,
and $t ,  \vec{r}$ are related to $t',  \vec{r} '$
by regular L-transformations [12].
Simple and seemingly obvious, (3)
doesn't however make happy match 
if any between relativistic QM on one hand, 
and EM-wave theory ($m_0 = 0$),
or classical QM ( $ |p|   \ll  m_0 c$ ) -- on the other hand,
for it brings about uneasy 
and unsettling ramifications.
In particular, an appropriately constructed
probability for particles with 
$ {E} > 0$ becomes negative 
for antiparticles, i. e. at $ {E}  < 0$.
Even at $ {E}  > 0$,
for such simple phenomenon as Doppler effect,
the solution (3) makes a wrong prediction
about the amplitude of wave,
as well as about non-conservation 
of the probability integral
for a limited wave-packet under frame-to-frame transfer 
(see below).

To address those issues,
as $\bf {a \ first \ step}$ we show that 
the KG-function $\Psi$ due to stipulation (3) is
physically inadequate, which can be fixed
by constructing a next-order 4-object,
an ansatz 4-vector (i. e. L-$covariant$), 
$\tilde{X} ( t , \vec{r} ) = ( \psi , \vec{X} )$,
based on 4-scalar $\Psi$, by using again a 4-gradient operator,
$\hat{{\tilde{\mathcal{D}}}}$ (1),
acting upon $\Psi$, so that
%
\begin{equation}
\tilde{X} = i \Lambda \hat{\tilde{\mathcal{D}}} \tilde{\Psi} ; \ \ or \ \
\psi = i \Lambda \partial \Psi / c \partial t ;  \ \ \ 
\vec{X} = - i \Lambda \hat{\vec{\nabla} } \ \Psi 
\tag{4}
\end{equation}
where $\Lambda = inv $
is an arbitrary constant and
is chosen by us as
$\Lambda = | \hbar | / m_0 c^2 > 0$
to match the dimensions of $\psi$ and $\Psi$,
and their non-relativistic magnitude.
(The use of $| \hbar |$ $vs$ $\hbar$
is intentional, see below.)
Since both $\hat{\tilde{\mathcal{D}}}$
and $\tilde{\Psi}$ are 4-objects,
$\tilde{X}$ is a 4-vector;
this can also be directly verified
by checking out that its components
transform as 4-radius-vector:
$\psi = \gamma   {\bf (} \psi ' + \beta X'_{\parallel} {\bf )}$,
$X_{\parallel} = \gamma   {\bf (} X'_{\parallel} + \beta  \psi ' {\bf )}$,
$\vec{X}_{\perp} = \vec{X} '_{\perp}$,
where $\beta = {\bf v} / c$,
and $\gamma =  $ $( 1 - \beta^2 )^{-1/2}$
with $\vec{{\bf v}}$ being the velocity
of moving frame in respect to lab-frame,
subscript $\parallel$ denotes
a vector component parallel to the velocity $\vec{{\bf v}}$,
and $\perp$ -- normal to it.

The scalar part, $\psi$, of $\tilde{X}$
is then what we are looking for,
a "true" wavefunction per se that
L-transformation-wise 
has the right properties, see below,
while an associated 3D-vector, $\vec{X} $,
is related to the momentum of the system.
$\psi$ is a solution of eq (3), same as $\Psi$,
with the difference being that
instead of the condition (3) on $\Psi$,
we have now a Lorentz gauge relation
between $\psi$ and $\vec{X}$ as
${\partial \vec{X} } / 
{ ( c \partial t ) } = -   \vec{\nabla}  \psi $.
Both $\psi$ and $\psi '$ can
be computed either from
$\tilde{\Psi} ( t , \vec{r} )$, if it is known --
for $\psi ( t , \vec{r} )$ from
(4), and for $\psi ' ( t' , \vec{r} ' )$ as
\begin{equation}
{ \psi ' ( t' , \vec{r} ' )} = 
i \Lambda {\partial \Psi {\bf \big[} \ t ( t' , \vec{r} ' ) , \
\vec{r}  ( t' , \vec{r} ' ) \ {\bf \big]} } / { c \partial t' }
\tag{5}
\end{equation}
or, if $\psi ( t , \vec{r} )$ is known, we have
$ \Psi ( t , \vec{r} ) = -  i (c  / \Lambda ) $
$\int \psi ( t , $ $\vec{r} ) \partial t$,
where $\int \psi \partial t$ stands for integration
over $\psi$ as an $explicit$ function of $t$,
and $\psi ' ( t' , \vec{r} ' )$
is evaluated $via$ (5).
At that, $\Psi$ (3), still is important as a "base" solution.

To demonstrate properties of wavefunction $\psi$,
we consider a plane wave propagation
with an energy $E$,
momentum $p$, and
amplitude $A$ of $\Psi$-function,
in the $x$-axis $||$ $\vec{{\bf v}}$.
Eq. (2) reduces then to 
a 1D-equation in space, whose solutions
$\Psi$ (3) and $\psi$ (4), (5) respectively
are as
\begin{equation}
\Psi = A e^{{i} ( p x - E t ) / \hbar } , \ \ \
\Psi ' = A e^{{i} ( p' x' - E' t' )  / \hbar  } 
\ \ \ 
\tag{6}
\end{equation}
with $E^2 = (pc)^2 + E_0^2$, 
$E_0 = m_0 c^2$, and
\begin{equation}
\psi = \Lambda ( {E} / {\hbar} ) \Psi , \ \ \
\psi ' = \Lambda (  {E'} / {\hbar} ) \Psi ' 
\tag{7}
\end{equation}
Since $E/c$ and $\vec{p}$ form a 4-vector,
their L-transformation is
$E = \gamma ( E' + \beta p' c )$,
$p = \gamma ( p' + \beta E' /c )$,
which entails a relativistic
Doppler effect $E (E' )$ as
\begin{equation}
E = \gamma  \Big( E' + 
\beta \sqrt { E'^2 - E_0^2}   \ \Big) \ \ \ with \ \ \
\ \ \
E_0 = m_0 c^2 .
\tag{8}
\end{equation}
In non-relativistic case, whereby
$\chi , \ \beta^2 \ll 1 $, with $\chi  \equiv E / E_0 - 1$,
we have
$\sqrt {\chi} = \sqrt { \chi ' } + \beta / \sqrt 2 $. 
For $m_0 = E_0 = 0$, (8) reduces to a familiar EM Doppler effect:
\begin{equation}
D_{EM} \equiv E / E' =
{\psi} / {\psi '} 
= \sqrt { {( 1 + \beta ) } / { ( 1 - \beta ) } }  
\tag{9}
\end{equation} %
reflecting the change of both frequency of photon,
and of the amplitude, $\psi / {\psi ' }$, 
of wavefunction (due to the change of photon count per second),
exactly the same as for classical electrodynamics.
The difference between KG solution 
$\Psi$ (6) and wavefunction $\psi$ (7) is that
the former one does not change its amplitude
under transformation between  lab and moving frames.
The same is true for $m_0 \ne 0$ 
since the relationship $\psi / {\psi ' } = E / E'$
remains valid here, which follows from (7) having in mind
that the phase $\theta = ( px - Et ) / \hbar = 
( p' x' - E' t' ) / \hbar $ is L-invariant.

Contemplating why the stipulation (3)
is wrong in application to a true wavefunction 
we want to point out that to require it to be  
L-$invariant$ is an undue overreach.
The relativity principle postulates
that physical phenomena occur the same way in any 
inertially moving frame, so 
the operators of the equations
of motion must be L-invariant, 
but the motion itself is L-$covariant$.
A good example is Maxwell wave equation 
whose operators
are L-invariant, while the fields 
$\mathcal{E}$ and $B$ are L-covariant [13].
A 4-scalar (3) by its definition is L-invariant;
a minimum L-covariant solution set has to be a 4-vector.

Independently of magnitude,
the phase $\theta = ( px - Et ) / \hbar = $ $inv$ 
in plane wave solutions (6) and (7) provides
a connection to both classical relativistic mechanics
on the one hand, and wave nature of QM,
on the other, $via$ a group, $\beta_{gr} \equiv { d E } / { c d p }$,
and phase $\beta_{phs} \equiv E / cp = \beta_{gr}^{-1}$,
velocities (i. e. $\beta_{gr} \beta_{phs} = 1 $).
If a wavefunction in a moving frame $K'$
becomes immobile ($p ' = 0$) at some
velocity $\beta = \beta_{cl}$,
we have $\beta_{cl} = cp / E = \beta_{gr}$.
Thus, as expected, $\beta_{gr}$ coincides with
a classical relativistic velocity;
simple calculations show that the L-transformation
for $\beta_{gr}$ coincides with familiar
relativistic velocity transformation:
\begin{equation}
\beta '_{gr} = {( \beta_{gr} - \beta ) } 
/ {( 1 - \beta \beta_{gr} ) } ;     \ \ \
{\beta_{gr}}^2 ,  {\beta '_{gr}}^2  
\in  {\bf [} 0 , 1 {\bf ]}
\tag{10}
\end{equation}
The same formula is true 
for $\beta_{phs}$ reflecting the wave nature of QM,
i. e.  if subscript "$gr$" is replaced by
"$phs$", albeit now $\beta_{phs}^2 ,  {\beta '}_{phs}^2  
\in  {\bf [} 1 , \infty {\bf )}$.
For photons, 
$\beta_{gr}^2 = {\beta '}_{gr}^2 =$
$\beta_{phs}^2 = {\beta '}_{phs}^2 = 1$
as expected.

Our $\bf {next \ step}$ is to consistently
solve the problem of probability 
density $w ( t , \vec{r} )$,
especially  for negative energies,
which is a thorny issue for KG-equation and its solutions.
In non-relativistic QM,
whereby $w =  | \psi |^2$ due to Max Born interpretation,
both integrals $W_{\psi} = \int w dV = <\psi  | \psi>$
and $W_{\Psi} =$ $ < \Psi  | \Psi >$ over the entire space 
are time-invariant full probabilities (when normalized);
but they are not L-invariant.
(In fact, for $any$ solution $\psi_{gen}$ of eq. (2),
$<\psi_{gen} | \psi_{gen}>$ is not L-invariant.)  
An analytically solvable example to illustrate this
is an ultimate relativistic case, EM-wave, $m_0 = 0$.
By choosing a traveling in the $x$-axis KG-solution
$\Psi ( t, x ) = F ( x - ct ) $ 
with an arbitrary real profile $F ( s )$ 
we have $W_{\Psi} =$ $ \int_{{-} \infty}^{\infty}  F^2 ( s ) ds$
whereas it can be shown that $W'_{\Psi} = W_{\Psi} D_{EM}$
where $D_{EM}$ is as in (9), 
so that $W_{\Psi}$ is not L-invariant.
A similar integral for $\psi \propto$ $  \partial \Psi / \partial t $,
$W_{\psi} = \int_{{-} \infty}^{\infty}
[{d F ( s  )} / {d s } ]^2 ds $
is not L-invariant either, since 
$W'_{\psi} = W_{\psi} / D_{EM}$.
This suggests that a "true" relativistic probability $W_{RL}$
is found "in-between" and is associated with 
the product $\Psi^* \cdot \psi $.
Decomposing a general solution of (2)
into plane wave solutions orthogonal to each other
(e. g. if $\vec{p}_1 \ || \ \vec{p}_2$, then 
$< \Psi ( \vec{p}_1 )  | $ $ \Psi ( \vec{p}_2 ) > 
\ \propto  \delta  ( p_1 - p_2 ) $,
and stipulating that 
(a) probability density, $w ( t, \vec{r} )$
has to be a non-imaginary
combination of bilinear products
of wavefunctions or their derivatives,
(b) that $w$ can be reduced to that of non-relativistic
QM, i. e. $w =  | \psi |^2$ at 
$| {E} | / E_0  -$ $ 1 \ll 1$,
and that $W = \int w dV$ has to be both (c)
L-invariant  and (d) $t$-invariant,
$d W / d t = 0$, we arrive at 
\begin{equation}
w = Re (  \Psi^* \cdot \psi ) ;    \ \ \
W = Re ( < \Psi  | \psi > ) ;
\tag{11}
\end{equation}
which (after normalization) makes a probability 
definition in general relativistic case [14].
Notice that while requiring that
$w$ is good in non-relativistic limit,
we assumed positive energy, $E > 0$.
However, as is well known, 
problems mount when we move to $E < 0$, 
which should corresponds to antiparticles.
As an example, evaluating the probability
density $w$ in (11), 
for a plane wave (6), (7),
and recalling that
$\Lambda = | \hbar | / E_0 $, eq. (4),
we obtain 
\begin{equation}
w = \Psi^* \psi = \Psi \psi^* =
{\frac{\hbar}{| \hbar | }}
\frac{E_0}{E} | \psi |^2 =
{\frac{| \hbar |}{ \hbar }}
\frac{E}{E_0} | \Psi |^2
\tag{12}
\end{equation}
i. e. $w < 0$ when $E < 0$, if $\hbar > 0$.
Let us make it very clear now
what KG-equation (2) and its solutions
with $E < 0$ are all about.
By writing $E^2 = (pc)^2 +E_0 ^2$ as in (2),
we in fact expanded the Einstein's
formula far beyond its original form 
$E = \sqrt { (pc)^2 + E_0^2}$
(while presuming $E > 0$, since 
at $p = 0$, $E = E_0 > 0$),
and thus allowed the energy $E$,
including the rest energy, $ {E}  =  - E_0$,
to be negative.
How come that relativistic QM so 
dramatically overturned the rock-solid
foundation of classical relativistic physics
without paying an equally fundamental
"passage fee" for that?
Strikingly obvious, and at that
no less strikingly overlooked/unexpected
is a simple proposition:
antiparticles have the sign of 
a singularly fundamental quantum 
parameter, $\hbar$,
reversed, hence the negative energies,
so the ratio $E / \hbar$ in (12)
remains invariant at the 
particle$\leftrightarrow$antiparticle 
transformation (PAT)
(see details below, eqns. (14), (15),
and related discussion).
With that $\hbar$-conjugation being
a passage rite at a gate between
the world and antiworld,
the classical physics is reversed too.
[Note that eq. (12) could now be rewritten
as $w = ( {E_0} / {|E|} ) | \psi | ^2 =
( |{E}| / {E_0} ) | \Psi | ^2$,
with $\psi / \Psi = {|E|} /  {E_0} $.
In non-relativistic
limit, $ |{E}| \rightarrow E_0$,
we have $\psi = \Psi$,
and $w = | \psi | ^2 = | \Psi | ^2$.]

All charged PAT pairs
are presumed to be comprised of oppositely charged
but otherwise similar particles. 
It can be shown, however, see below,
that aside from $\hbar$-conjugation,
other particle characteristics, 
in particular gravitational mass, 
reverse their sign at PAT too. 
This implies an "inverse" reaction of 
antiparticles, including antigravity, 
to any external force,
which greatly affects the properties
of neutral macro-antimatter.
Thus charge-0 particles have their 
well distinguishable antiparticles too.
(The antigravity between matter and antimatter
would have profound implications for cosmology, see below.)
All this makes antiparticles different
physical species that must be
described by $separate$ wavefunctions,
$\Psi_{\downarrow}$ (or $\psi_{\downarrow}$), $vs$ those
for regular particles $\Psi_{ \uparrow }$ (or $\psi_{ \uparrow }$),
with subscript $ \uparrow $ 
assigned to particles, and 
$\downarrow$ -- to antiparticles.
Thus two kinds of solutions 
($ \uparrow $ and $\downarrow$)
generated by the KG-equation (2) for $m_0 \ne 0$
and defined by the sign of $\hbar$,
$\hbar_{\downarrow} = - \hbar_{ \uparrow } $,
are unrelated to each other.
The respective probability densities, 
$w_{ \uparrow }$ and $w_{\downarrow}$ in (12), 
ascribed to different entities,
must be also unrelated,
and in general $unequal$, yet
always positively-defined.

As our $\bf {finall \ step}$,
we show that all the matter-field
interactions indeed reverse their sign
for PAT along with $\hbar$,
by fully incorporating them into 
expanded, or "dressed" [15] KG-equation.
To keep it L-invariant, we may use
only those that form
4-vectors (as e. g.  EM or gravitation potentials [16]).
Assigning a 4-potential to each one
of them, $q_j \tilde{\Phi}_j =
q_j ( \phi_j , \vec{A}_j )$,
where $\phi_j$ and $ \vec{A}_j$
are respectively scalar and vector potentials
of the $j$-th interaction, and $q_j$
are the "charges" of a particle quantifying
its reaction to related fields,
e. g. an electron charge $e$ or gravitational mass $m_G$;
other particle-field interactions, if any,
should be included here too.
(The dimension of potentials $\tilde{\Phi}_j$
is to be such that 
$q_j \phi_j$ has the same dimension as $ {E}  / c$,
and $q_j \vec{A}_j$ -- as $\vec{p}$).
All of them can be coupled to 4-momentum 
$\tilde{p} = ( { {E} } / c , {\vec{p}} )$, 
using a generalized 4-momentum $\tilde{P}
= \tilde{p} + \Sigma q_j  \tilde{\Phi}_j$, i.e.
%
\vspace {-.05in}
\begin{equation}
\tilde{P} = 
\big( \   {E}  / c + \Sigma q_j \phi_j ( \vec{r} , t ) / c ,
\
 \vec{p} + \Sigma q_j \vec{A}_j ( \vec{r} , t ) \ \big)
\tag{13}
\end{equation}
Translated into operator form using (1)
for $\hat{ {E} }$ and
$\hat{\vec{p}}$, the fully dressed
KG-equation reads then as
\begin{equation}
{\bf \big[} - {\bf (} i \hbar \partial / c \partial t + \Sigma
q_j \phi_j {\bf /} c {\bf )}^2 +
\notag
\end{equation}
\vspace {-.25in}
\begin{equation}
{\bf (} - i \hbar \vec{\nabla} +
\Sigma q_j \vec{A}_j {\bf )}^2
+ ( m_0 c )^2 {\bf \big]} \psi ( \vec{r} , t ) = 0
\tag{14}
\end{equation}
examination of which reveals that
in the presence of any external interactions
(i. e. $\Sigma q_j  \tilde{\Phi}_j \ne 0$)
{\it {eq. (14) is invariant to the simultaneous sign-reversal
of $\hbar$ (and related operators
$\hat{ {E} }$ and $\hat{\vec{p}}$),
as well as of ALL the charges, $q_j$},
\rm 
regardless of time\&space symmetries  
of $\phi_j$ and $\vec{A}_j$
so that a fundamental relationship proves to be true:
\begin{equation}
{\bf \big\{ } \hbar , e , m_G ..., 
\hat{ {E} } , \hat{\vec{p}} \ {\bf \big\} }_{\downarrow} =
- {\bf \big\{ } \hbar , e , m_G ..., 
\hat{ {E} } , \hat{\vec{p}} \ {\bf \big\} }_{\uparrow } ,
\tag{15}
\end{equation}
This makes a "fully reversible jacket" [17] (FRJ)
for the dressed KG-equation (14), and
constitutes a new invariance
to be satisfied for any relativistic wave equation.
(Note that because of the structure of 
(14), inertial mass $m_0$ is $not$ part of FRJ.)
The most important observation about (15) is that, 
\it due to $\hbar$-conjugation, 
it completely eliminates the relevance of
$CPT$-invariance in PAT theory [i. e. of
the invariance of (14) with respect to
sign reversal of 
(charge)+(spatial coordinates/parity)+(time)].
\rm
This may call for a fundamental
revision in many applications of $CPT$-symmetry.

An illuminating view on PAT
issue transpires from breaking all the
entrants in (14) into 
two subsets: $arena$ (time+space+inertia), and
$players$ (fields+interactions).
Without players, (14) reduces to (2),
which is trivially invariant to any combination of 
$t$, $\vec{r}$, or $m_0$ sign reversals.
Of major significance (if $m_0 \ne 0$)
however, is a players' supper-symmetry (15),
which completely ignores 
${\bf \big\{} t, \vec{r}, m_0 {\bf \big\}}$ arena,
making the PT-symmetry irrelevant
(including the Stueckelberg-Feynman
interpretation [18] of PAT)
and offering a simple and 
intuitively transparent view of PAT.
(Notice that antiparticle
gravitational mass $( m_G )_{\downarrow}$
sign is opposite to its inertial mass, $ m_0 > 0$.)
From math-viewpoint,
$\hbar$-conjugation is not isomorphic
to $\vec{r} \& t$ sign-reversal in general:
if $\phi_j$ and $\Phi_j$ have no such symmetry,
and $\hbar = | \hbar |$,
invariance of (14) does not hold.
There is little doubt that 
$\hbar$-conjugation principle would hold for other 
quantum relativistic theories of
antiparticles, including Dirac equation [5]
for spin-1/2, and Proca equation 
[19,3] for spin-1 particles.

When applied to averaged energy $<{E}>$
and temperature $T$, FRJ-symmetry (15) yields
$\{ K_B , T  ,  <{E}>   \}_{\downarrow} = -$
${\bf \{ } K_B , T  ,  $ $<{E}>   {\bf \} }_{\uparrow}$,
where $K_B$ is the Boltzmann constant.
However, the product $K_B T$ in
Maxwell-Boltzmann distribution,
as well as in
its Maxwell-J\"{u}ttner generalization
for relativistic gas [20,21], are FRJ-invariant.

For photons,
at $m_0 = q_j = 0$ in (14) $\hbar$-term cancels out;
what is left is an EM wave-equation, 
$\Box \psi = 0$.
A plane wave
$\psi \propto \exp [ i ( x - c t   ) \omega / c ]$
misses $\hbar$ and has no "anti"-solution,
so there are no antiphotons.
From the antiparticle's viewpoint though,
photons have negative (i. e. acceptable)
energy, $E_{\downarrow} = \hbar_{\downarrow} \omega =
- | \hbar | \omega$ [22].

At high enough energies and densities of matter and radiation, 
the KG-equation, which implies conservation of particle numbers,
does not hold anymore and
is to be replaced by quantum field theory [6,23],
to address annihilation/creation of pairs,
photon-electron scattering, in particular 
Klein-Nishina theory [24,23],
radiation pressure at very high 
particle energies [21], etc.

One can project KG-equation (14)
into non-relativistic QM.
By using a FRJ operator, $\hat{\kappa}$,
with $\kappa_{ \uparrow } = 1$, 
$\kappa_{\downarrow} =$ $ - 1$, 
and non-relativistic wavefunction,
$\psi_{NR} $ $( t , \vec{r} ) = \psi  ( t , \vec{r} ) \exp 
( i  \hat{\kappa} E_0  t / | \hbar | )$,
we have from (14) and (15) 
Schr\"{o}dinger-like equations 
for either species:
\begin{equation}
{\bf \big\{} \hat{\kappa} \big[ i | \hbar | 
{ \partial } / { \partial t } 
- ( U_{\Sigma} )_{ \uparrow } \big]
 +  { \hbar^2 \nabla^2 } / {2 m_0 } 
{\bf \big\}}  \psi_{NR} = 0 ;    
\tag{16}
\end{equation}
with $( U_{\Sigma} )_{ \uparrow } = \Sigma q_j \phi_j =
- ( U_{\Sigma} )_{\downarrow}$,
which is derived under assumption
$ | U_{\Sigma} |  / E_0$  ,   
$| {E}  / E_0 - \hat{\kappa} |  \ll 1$,
and with the term $ \Sigma q_j \vec{A}_j $ in (14)
neglected as compared to $\hbar \vec{\nabla}$
due to Lorentz gauge applied to $\phi_j$ and $\vec{A}_j$.
Thus again, the antiparticles may be viewed not as
purely relativistic creatures,
but rather as an extension of QM and classical
mechanics into the domain of negative energies
and related FRJ-symmetry.
In particular, non-relativistic classical dynamics 
is governed by a FRJ-expanded Newton laws
separately for particles and antiparticles:
\vspace {-.01in}
\begin{equation}
{d \vec{p}} / dt ( or \   m_0 {d \vec{{\bf v}}} / dt ) = - 
\hat{\kappa} \nabla ( U_{\Sigma} )_{ \uparrow }
\tag{17}
\end{equation}
Thus, particles and antiparticles
are accelerated in opposite directions
in any external potential force, 
regardless of its nature.
In particular, a non-relativistic Newton's law 
for an $e$-charged particle in
EM fields $\vec{\mathcal{E}} ( \vec{r} )$, $\vec{B} ( \vec{r} )$
and gravitational force $ m_0 \vec{g} ( \vec{r} )$ is:
%
\begin{equation}
{d \vec{{\bf v}}} / dt =
\hat{\kappa} {\bf {\big\{}}  - ( e_{ \uparrow } / m_0 )
( \vec{\mathcal{E}}  + {\bf {[}} \vec{\beta} \times \vec{B} {\bf {]}} )
+ \vec{g} {\bf {\big\}}}
\tag{18}
\end{equation}
[based on $e_{ \uparrow } = - e_{\downarrow}$, and 
$( m_G )_{ \uparrow } = m_0 = - ( m_G )_{\downarrow}$].  
Antiatoms or neutral antimatter
will be rising up in the field of Earth gravitation
provided there are no other forces.

There are experiments
on the discussed effects that may conceivably be conducted 
in the future with antiparticle trapping. 
As we've mentioned before, a detailed characterization
of true wavefuntions (for either particles or antiparticles)
can be done by using high-precision
multi-interferometry in "quantum
carpets" [10] in weakly-relativistic case [11].
The most direct experiment on anti-gravity
could be monitoring 
atom/antiatom dynamics due to earth gravitation
in a trap [25] with suddenly removed trapping fields.
Another way to go might be to try and observe 
low-energy collisions between the beams of 
atoms and antiatoms that should repulse each other.
Other group of experiments may pertain to
observation of weak-relativistic cyclotron hysteresis [8], 
of separately excited proton and antiproton
(or electron, similarly to [8], and positron).
Trapped neutral atoms and 
antiatoms (or respective ions),
can be engaged in large self-sustained
motional oscillations [26,27]
due to the Doppler-affected radiation pressure
by a laser, blue-detuned from an 
absorption line of a particle;
huge hysteresises [26] of excited motion should
have their threshold and spectral characteristics
dependent on the sign of $\hbar$.
However, more detailed study of these effects
are to be done in near future.

Antigravity between matter and antimatter
is of no consequence in high-energy
interactions,
as it can be neglected 
compared to other forces.
However, if validated, it may prove to be 
a tremendous new factor in
the evolution of universe.
A possible scenario may be envisioned as this:
Following formation of baryons and antibaryons
in a due course of after-big-bang evolution,
and their consequent mass-annihilation,
some of them had survived in roughly 
$equal$ $numbers$ -- and $not$ $baryons$ $alone$
-- as opposed to widely accepted
baryogenesis model [28] of $only$ baryons remaining
due to their initial excess.
(That awkward model/hypothesis was invoked to
explain the preponderance of regular matter $vs$
anti-matter in our universe.)
Since the expansion at that period was fast,
the rate of further annihilation 
got greatly diminished once the densities
of both species got sufficiently low --
yet still almost equal.

Later on, after formation of electrons and positrons,
and then neutral matter and antimatter
as the universe cooled, the gravitation between
similar species and antigravitation between different ones,
became a major factor in further evolution.
The interplay of attraction and repulsion forces
lead to spatial instability promoting
separation of different species
and clustering of similar ones,
which in turn brought about a phase transition
ending up in formation of comb-like multitude of separated 
"checkered sub-universe" domains, 
with each of them consisting of
either almost pure matter or antimatter [29].
This may explain the preponderance of regular
matter over antimatter, 
on one hand, and apparent absence
of antigravity in our neck of the universe,
on the other hand,
as two facets of the same FRJ-phenomenon.

Closer to the "cool" low-density state of universe, 
those domains may begin to inter-penetrate and diffuse,
which may result in faster acceleration of their expansion 
due to reduced (on average) gravitation
in each initially homogeneous domain.
This in turn may substantially 
contribute to (if not fully explain)
the dark energy phenomenon.

In conclusion, by using "true"
KG-wavefunctions based on 4-vector
Lorentz-covariant solutions,
we predicted antiparticles 
whose Planck constant, $\hbar$, has its sign reversed.
This makes a root reason for negative energies and
"everything-conjugated" symmetries,
whereby all the parameters
related to interaction of particles
with external fields reverse
their sign for antiparticles,
including electrical charge and gravitational mass
(i. e. antimatter should exhibit 
anti-gravitation with respect to a regular matter).
The related FRJ-invariance greatly
simplifies the theory of
particle$\leftrightarrow$antiparticle
transformation by fully replacing $CPT$-invariance.

\begin{acknowledgments}
The author appreciates greatly helpful
comments by anonymous referees.
He thanks Prof. B. Ya. Zeldovich
for thoughtful comments in
the early stages of work on this paper, and
Prof. J. Fajans for illuminating 
outlook on the state of research by ALPHA collaboration in 
the antiHydrogen experiment at CERN.
He is also grateful to Profs.  
I. Averbukh, G. Kurizki, E. Pollak, and Y. Prior, 
of Weizmann Inst. of Science, 
for their kind hospitality
during his stay as a visiting researcher there
in the Falls of 2016 and 2017,
with special thanks to E. Pollak,
who organized a seminar on the subject
of this paper, and G. Kurizki
for critical reading of the original manuscript.
\end{acknowledgments}
%


\begin{thebibliography}{99}
%
\bibitem{cit1}
O. Klein,
Z. Phys., $\bf {37}$, 895 (1926) and $\bf {41}$, 407 (1927);
W. Gordon, 
Z. Phys., $\bf {40}$, 117 (1926).
Other authors of related work
in the same period were V. Fock, J. Kudar, 
T. de Donder, F.H. van den Dungen, 
and L. de Broglie.
%
\bibitem{cit2} 
W. Pauli and V. Weisskopf, 
Helv. Phys. Acta , $\bf {7}$, 709 (1934).
%
\bibitem{cit3}
W. Greiner,
Relativistic Quantum Mechanics: Wave Equations,
3-rd ed., Springer-Verlag, New  York (2000);
S. S. Schweber,
An Introduction to Relativistic Quantum Field Theory,
2-nd ed., Harper \& Raw, New York (1962).
%
\bibitem{cit4}
P.A.M. Dirac, Proc. R. Soc. (London),
$\bf {A117}$, 610 (1928), $ibid$.
$\bf {A118}$, 351 (1928);
Principles of Quantum Mechanics
4th ed., Clarendon Press, Oxford (1981).
%
\bibitem{cit5}
K. Gottfried and W.F. Weisskopf,
Concepts of Particle Physics,
Oxford Univ. Press, New York,
v. 1 (1984) and 2 (1986).
%
\bibitem{cit6}
S. Weinberg,
The Quantum Theory of Fields,
v. 1, Cambridge Univ.Press,  New York (1995)
%
\bibitem{cit7} 
P. Luchini and H. Motz, 
Undulators and free-electron lasers,
 Oxford Univ. Press, Oxford (1990).
%
\bibitem{cit8}
Weakly relativistic hysteretic cyclotron resonance 
was predicted in A.E. Kaplan, Phys. Rev. Letts.,
{\bf {48}}, 138 (1982),
and experimentally observed in
G. Gabrielse, H. Dehmelt, and W. Kells,
Phys. Rev. Lett. {\bf {54}}, 537 (1985).
%
\bibitem{cit9}
A. E. Kaplan and Y. J. Ding, 
IEEE J. Quant. Electr., {\bf {24}}, 1470 (1988).
%
\bibitem{cit10}
A. E. Kaplan, P. Stifter, K.A.H. van Leeuwen, 
W. E. Lamb, Jr. and W. P. Schleich, 
Physica Scripta T, {\bf { 76}}, 93 (1998);
A. E. Kaplan, I. Marzoli, W. E. Lamb Jr., 
and W. P. Schleich, 
Phys. Rev. A, {\bf { 61}}, 032101 (2000).
%
\bibitem{cit11}
I. Marzoli, A. E. Kaplan, F. Saif, and W. P. Schleich, 
Fortschritte der Physik, {\bf { 56}}, 967 (2008) 
%
\bibitem{cit12}
It is a not uncommon implication or direct statement in the literature
that $all$ the solutions of (2) have the same property as in (3),
which is not true, see eq (4) and on.
In fact, eq. (3) is a choice stipulation/limitation
imposed on those solutions, 
and not their general property.
%
\bibitem{cit13}
L.D. Landau and E.M. Lifshitz,
The Classical Theory of Fields,
4-th ed., Butterworth, Oxword (1994).
%
\bibitem{cit14}
In fact, eq (11) is consistent with 
the results obtained
$via$ current density
calculations only at $E > 0$, see e. g. [3]
(considering different definitions for $\psi$ and $\Psi$).
%
\bibitem{cit15}
Dressed states (i. e. quantum effects
due to perturbations in "base"
system driven by external fields)
gained broad recognition
and use in atomic physics and
quantum optics, see e. g.
C. Cohen-Tannoudji, G. Grynberg and J. Dupont-Roc,
Photons and Atoms: Introduction to Quantum
Electrodynamics, Wiley, New York (1997).
M.O. Scully and M. S. Zubairy,
Quantum Optics, Cambridge Univ. Press, Cambridge (1997);
P. Meystre, Atom Optics, Springer-Verlag, New York (2001)
%
\bibitem{cit16}
We do not discuss general relativity here,
so the gravitation potential
is assumed sufficiently weak.
%
\bibitem{cit17}
The author respectfully acknowledges a
connotation to the Stanley Kubrick's
"Full Metal Jacket" movie title.
%
\bibitem{cit18}
E. Stueckelberg, Helv. Phys. Acta $\bf {14}$, 322 (1941);
R.P. Feynman, 
Rev. Modern Physics, $\bf {20}$, 367 (1948). 
%
\bibitem{cit19}
A. Proca, J. Phys. Radium, $\bf {7}$, 347 (1936),
C. R. Acad. Sci. Paris, $\bf{202}$, 1490 (1936).
%
\bibitem{cit20}
F. J\"{u}ttner, Ann. Physik, $\bf {339}$, 856 (1911);
J. L. Synge, The Relativistic Gas, North-Holland, 1957.
%
\bibitem{cit21}
A.E. Kaplan, J. Phys. B, {\bf {48}}, 165001 (2015).
%
\bibitem{cit22}
The analogy of negative energy
in application to interaction
between photons and matter is common in 
quantum electronics and nonlinear optics,
where it is used in the theory
of difference frequency
generation, stimulated Raman scattering,
four-wave mixing, etc.,
N. Bloembergen,
Nonlinear Optics, W. A. Benjamin Publ., New York, 1965;
Y. R. Shen,
The Principles of Nonlinear Optics,
J. Wiley, New York, 1984;
V.S. Butylkin, A.E. Kaplan, Y.G. Khronopulo,
and E. I. Yakubovich, Resonant Nonlinear Interactions
of Light with Matter, Springer-Verlag, New York, 1989
(translated and updated from original, Nauka, Moscow, 1977).
%
\bibitem{cit23}
V.B. Beresteckii, E.M. Lifshitz, and L.P. Pitaevskii,
Quantum Electrodynamics,
2-nd ed., Pergamon, Oxword, 1982.
%
\bibitem{cit24}
O. Klein and Y. Nishina,
Z. Phys. $\bf {52}$, 853 and 869 (1929).
%
\bibitem{cit25}
Simulations for the planned experiments at CERN, 
A.E. Charman, C. Amole, M.D. Ashkezari, et al.
(ALPHA Collaboration group),
Nature Comms., $\bf {54}$, 1785 (2013),
with antiHydrogen atoms
to be produced and trapped in a magnetic field
exceeding 1 T, and escaping atoms
detected as they annihilate at the trap wall,
showed that the window for the ratio
$m_G / m_0$ at this point is estimated
to still greatly exceed $\pm 1$,
yet has strong potential for enhancement.
%
\bibitem{cit26}
A. E. Kaplan,
Optics Express, $\bf {17}$, 10035 (2009)
%
\bibitem{cit27}
K. Vahala, M. Herrmann, S. Kn\"{u}nz, V. Batteiger,
G. Saathoff, T. W. H\"{a}nsch and Th. Udem,
Nature Physics, $\bf { 5}$, 682 (2009).
%
\bibitem{cit28}
S. Weinberg, Cosmology, Oxford Univ. Press, Oxford, 2008;
P.J.E. Peebles,
Principles of Physical Cosmology,
Princeton, Princeton Univ. Press, 1993;
Y.B. Zel'dovich and I.D. Novikov,
Relativistic Astrophysics, v. 2,
Univ. Chicago Press, Chicago, 1983.
%
\bibitem{cit29}
The obvious requirement 
for such domains is that each of
their 2D facets/faces must be a divider
between alternating matter and anti-matter,
with each 1D edge being thus an intersection
of even number of faces.
The simplest topological 
structure of those domains
would be then basic 3D comb of either 
tetrahedron-like or cubic-like cells.
Each of those is a topological 3-sphere;
of further interest might be 
the cells of a higher-order topological
structure, e. g. formed 
by toruses or Klein bottles.
%
\end{thebibliography}
\end{document}